\documentclass[sigconf,authorversion]{acmart}

\renewcommand\footnotetextcopyrightpermission[1]{}

\AtBeginDocument{%
  \providecommand\BibTeX{{%
    \normalfont B\kern-0.5em{\scshape i\kern-0.25em b}\kern-0.8em\TeX}}}

\setcopyright{acmlicensed}
\copyrightyear{2018}
\acmYear{2018}
\acmDOI{XXXXXXX.XXXXXXX}

\acmConference[Conference acronym 'XX]{Make sure to enter the correct
  conference title from your rights confirmation emai}{June 03--05,
  2018}{Woodstock, NY}
\acmISBN{978-1-4503-XXXX-X/18/06}

\usepackage{multirow}
\usepackage{makecell}

\usepackage{graphicx}
\usepackage{subfigure}

\definecolor{cadmiumgreen}{rgb}{0.0, 0.42, 0.24}
\definecolor{carrotorange}{rgb}{0.93, 0.57, 0.13}

\begin{document}

\title{TRAWL: External Knowledge-Enhanced Recommendation with LLM Assistance}

\author{Weiqing Luo}
\email{wqluo@smail.nju.edu.cn}
\affiliation{%
  \institution{State Key Laboratory for Novel Software Technology, Nanjing University}
  \institution{WeChat, Tencent}
  \country{China}
}

\author{Chonggang Song}
\email{jerrycgsong@tencent.com}
\affiliation{%
  \institution{WeChat, Tencent}
  \country{China}
}

\author{Lingling Yi}
\email{chrisyi@tencent.com}
\affiliation{%
  \institution{WeChat, Tencent}
  \country{China}
}

\author{Gong Cheng}
\email{gcheng@nju.edu.cn}
\affiliation{%
  \institution{State Key Laboratory for Novel Software Technology, Nanjing University}
  \country{China}
}

\begin{abstract}

Combining semantic information with behavioral data is a crucial research area in recommender systems. A promising approach involves leveraging  external knowledge to enrich behavioral-based recommender systems with abundant semantic information. However, this approach faces two primary challenges: denoising raw external knowledge and adapting semantic representations.
To address these challenges, we propose an External Knowledge-Enhanced Recommendation method with LLM Assistance (TRAWL). This method utilizes large language models (LLMs) to extract relevant recommendation knowledge from raw external data and employs a contrastive learning strategy for adapter training. Experiments on public datasets and real-world online recommender systems validate the effectiveness of our approach.

\end{abstract}

\maketitle

\section{Introduction}
\label{Sec:Intro}

Recommendation systems contain a wealth of semantic information, such as product names and descriptions, and user profiles. This semantic information can naturally be linked to open-world knowledge. For example, it is a common practice to introduce prior external knowledge into recommendation systems through entity linking techniques~\cite{pham2014recommendation} and hence improve recommendation performance with the incorporation of supplementary information.

The current mainstream approach in recommendation systems is to model user and item IDs, leveraging behavioral collaborative signals to enhance recommendation performance~\cite{das2017survey}. 
ID embeddings are primarily designed to encapsulate the behavioral correlations exhibited by user-item interactions. However,  these embeddings frequently fail to  capture the content relevance inherent to the items themselves. This limitation is particularly pronounced in scenarios such as the cold start problem and in the context of similarity-based recommendations
How to effectively utilize the aforementioned prior external knowledge to complement the behavioral information in recommendation systems and thereby improve performance has not been thoroughly studied~\cite{guo2020survey}.

Intuitively, leveraging external knowledge involves a two-stage process: firstly, it requires the extraction of effective representations of the semantic information. Subsequently, these representations must be efficiently integrated into downstream recommendation algorithms. Each stage poses substantial challenges.

\textbf{Challenge 1.} \emph{Raw external knowledge is not inherently tailored to recommender systems and necessitates processing to become truly beneficial for recommendation.}

External knowledge, though readily obtainable through prevalent semantic information in common recommender systems, poses significant utilization challenges, which can be encapsulated into two primary gaps. 

The first gap arises from the noise present in the raw external knowledge, which, although not necessarily detrimental to recommendation performance, requires careful handling. For instance, in movie recommender systems, semantic information is available for both items and users. 
For a specific movie, it is straightforward to acquire world knowledge from a knowledge base. Similarly, for a particular user, in addition to his (or her) basic profile, the external knowledge of movies in their viewing history is a valuable resource for recommendation. 
However, the raw external knowledge for a movie is often extensive, with useful information dispersed throughout, necessitating extraction and summarization. %

The second gap pertains to knowledge that is implicitly beneficial for recommendation, which involves transforming knowledge that is outside the immediate recommendation context through common-sense inference. 
For instance, encoding the numerical age of a user under 18 directly into the language model offers limited utility to the recommendation system. However, applying common-sense inference can deduce that this user’s preferences are more likely to include content relevant to teenagers, such as fantasy films, thereby significantly improving recommendation accuracy.

\emph{In essence, extracting truly beneficial recommendation knowledge from raw external knowledge necessitates that the recommender system develop capabilities for information filtering and common-sense inference.}

\textbf{Challenge 2.} \emph{Adapting recommendation-related content extracted from external knowledge from the semantic space to the recommendation space is an immensely challenging task.}

Even when raw external knowledge has been processed into useful recommendation knowledge and encoded into semantic representations by natural language processing tools, a gap remains between these semantic representations and behavioral representations. This disparity complicates the integration and utilization of both types of information within the recommender system. \emph{Consequently, to effectively combine external knowledge with behavioral information, the recommender system must acquire the capability to adapt semantic representations to the recommendation space.}

\textbf{Our Work}

To address the aforementioned challenges, we propose our recommendation framework, TRAWL (Ex\underline{t}ernal knowledge-enhanced \underline{R}ecommend\underline{a}tion \underline{w}ith \underline{L}LM).

For Challenge~1, TRAWL addresses the two identified gaps by using appropriate prompts to guide a large language model (LLM). 
First, TRAWL directs the LLM to concentrate on the key recommendation-related elements of the raw external knowledge while filtering out noisy parts, leveraging the LLM's advanced text summarization capabilities. Second, utilizing the LLM's common-sense inference abilities, TRAWL facilitates the integration of world knowledge with recommendation knowledge, thereby generating more useful semantic representations for the downstream recommender system. Different from traditional approaches employed for the generation of textual summaries, LLMs exhibit a distinct advantage in their capacity  in synthesizing summaries that are not only concise but also significantly enriched with pertinent information, contingent upon the provision of specific cues or prompts.
TRAWL generates recommendation knowledge for both the user and item sides, ensuring alignment between them through uniform key factors in the prompts.

For Challenge~2, TRAWL integrates domain knowledge (i.e., user-item behavioral information) as supervisory signals to train the adaptation module. We employ multi-task learning to address the recommendation and adaptation tasks concurrently. Specifically, in addition to the original loss of the recommendation network, we introduce a novel auxiliary loss function to train the adaptation module. This module is designed to transform the recommendation knowledge generated by the LLM into information beneficial for recommendations. By incorporating domain knowledge, we train the adaptor to embed behavioral information into the semantic representations, ensuring that users and items with similar interactions produce similar representations.

We conduct extensive experiments to validate the effectiveness and generalizability of our framework. In addition to experiments on public academic recommendation datasets, we deploy TRAWL in a real-world recommendation platform on WeChat to validate its effectiveness.

Our contributions are summarized as follows:
 \begin{itemize}

\vspace{-0.5cm}
\item[1.] We propose a pioneering framework, TRAWL, which utilizes LLM to extract recommendation knowledge from raw external knowledge. TRAWL is versatile and not limited by the choice of recommendation algorithm, making it easily integrable with any recommendation system.

\item[2.] We design a contrastive learning approach, coupled with multi-task learning, to train an adaptor that bridges the gap between semantic information and behavioral information.

\item[3.] We validate the framework's effectiveness through comprehensive experiments. To further validate the effectiveness of our framework, we also present an industrial deployment on a large-scale online platform.   %
\end{itemize}

\section{Related Work}
\label{sec:related-work}

In this section, we introduce the related studies on using external knowledge for recommendation.

\subsection{External Knowledge Denoising and Extraction}
Raw external knowledge typically contains a significant amount of noise. To mitigate its negative influence, methods can be pursued in two different directions: denoising, which involves filtering out the useless parts, and extraction, which focuses on identifying and extracting the useful parts. Additionally, the form in which the external knowledge is organized also influences the method design.

\subsubsection{External Knowledge in Graph Form}
\label{subsubsec:related-work-external-klg-in-graph-form}

Knowledge graphs (KGs) are valuable resources for improving the performance of recommender systems, and numerous studies have focused on addressing the noise in KGs to obtain more useful semantic representations. DiffKG~\cite{jiang2024diffkg} effectively integrates knowledge graph information through a knowledge graph diffusion model and a collaborative knowledge graph convolution mechanism. This approach mitigates the negative impact of noise in the knowledge graph, thereby enhancing the performance and robustness of recommender systems. Similarly, to address noise issues in both knowledge graphs and user interaction data, KRDN~\cite{zhu2023krdn} employs adaptive knowledge refinement and contrastive denoising learning to effectively extract valuable information from noisy data, providing more accurate and robust recommendations. 

The aforementioned works focus on denoising external knowledge in KGs. \emph{In contrast, this paper assumes external knowledge is in a freely organized text form (e.g., an introductory article about a particular item) and does not require it to be in a structured form.} We will not refer to KG-based methods in the following sections.

\subsubsection{External Knowledge in Text Form}

In addition to graph-formed external knowledge, a substantial amount of knowledge is unstructured, rendering the methods in Section~\ref{subsubsec:related-work-external-klg-in-graph-form} inapplicable. A straightforward natural language processing technique to denoise (or extract) useful recommendation knowledge from raw external knowledge is summarization, a well-established research area prior to the era of large language models~\cite{el2021text_sum_survey}. Basic summarization methods, such as PEGASUS~\cite{zhang2020pegasus}, achieve high-quality text summarization by training a Transformer model with a self-supervised pre-training objective called Gap Sentences Generation (GSG), which involves generating gap sentences within a document. A sub-domain task, topic-focused text summarization, generates summaries guided by a user-provided topic~\cite{tang2009topic_focus}. Representative works include CTRLsum~\cite{he2020ctrlsum}, which introduces control tokens during training and inference, allowing users to interact with the summarization system using keywords or descriptive prompts, thereby achieving multifaceted control over the generated summaries.

\emph{As introduced in Section~\ref{Sec:Intro}, extracting recommendation knowledge from raw external knowledge in text form is a challenging task, akin to a more complex version of topic-focused text summarization. In our work, we choose to use LLM to complete this task, and we will report comparative experiment results between using LLM and the aforementioned text summarization methods in Section~\ref{sec:exp-results}}.

\subsection{LLM-generated External Knowledge for Recommendation}

Large language models (LLMs) have been demonstrated to contain extensive open-world knowledge~\cite{zhao2023llmsurvey}, leading to research on using LLMs to generate external knowledge for downstream recommender systems. For example, KAR~\cite{xi2023towards} leverages factual knowledge stored in LLMs to enhance movie and book recommendations. LLMRec~\cite{wei2023llmrec} employs LLMs to enhance user and item profiles and constructs pseudo-supervised data using LLMs to address data sparsity issues in the recommendation domain. Similarly, ONCE~\cite{liu2024once} uses prompt techniques to enable closed-source large models to generate new item content features and improve user profiles, thereby enhancing the performance of recommendation systems.

The differences between these works and ours are significant: they directly utilize LLMs to generate external knowledge, \emph{whereas we employ LLMs in an auxiliary role to extract recommendation knowledge from existing external sources. By generating knowledge from real-world information, we effectively mitigate the negative impact of the well-documented hallucination problem~\cite{zhao2023llmsurvey}, thereby enhancing the reliability of the generated knowledge.}

\subsection{Semantic Representation Adaptation for Recommendation}

As introduced in Section~\ref{Sec:Intro}, the disparity between semantic representations and behavioral information leads to suboptimal performance when utilizing external knowledge. Previous studies have attempted to bridge this gap through various adaptation methods. For instance, UniSRec~\cite{hou2022unisrec} trains a Mix-of-Experts (MoE) network to adapt textual representations of items into a universal representation across different domains. Similarly, KAR~\cite{xi2023towards} and SSNA~\cite{peng2023ssna} employ MoE adapters to adjust LLM outputs to fit the recommendation space. 

In this work, we also utilize an adapter network to address this gap. \emph{However, instead of complicating the design of the adapter network, we  propose a contrastive learning method coupled with multi-task learning to bridge the gap. This approach is orthogonal to previous works and can be combined with them.}

\section{Method}
\label{sec:method}

This section provides a comprehensive introduction of the TRAWL framework. Initially, we present an overview of the framework. Subsequently, we delve into the design specifics of individual modules in the following subsections.

\begin{figure*}
\centering
\includegraphics[width=1.75\columnwidth]{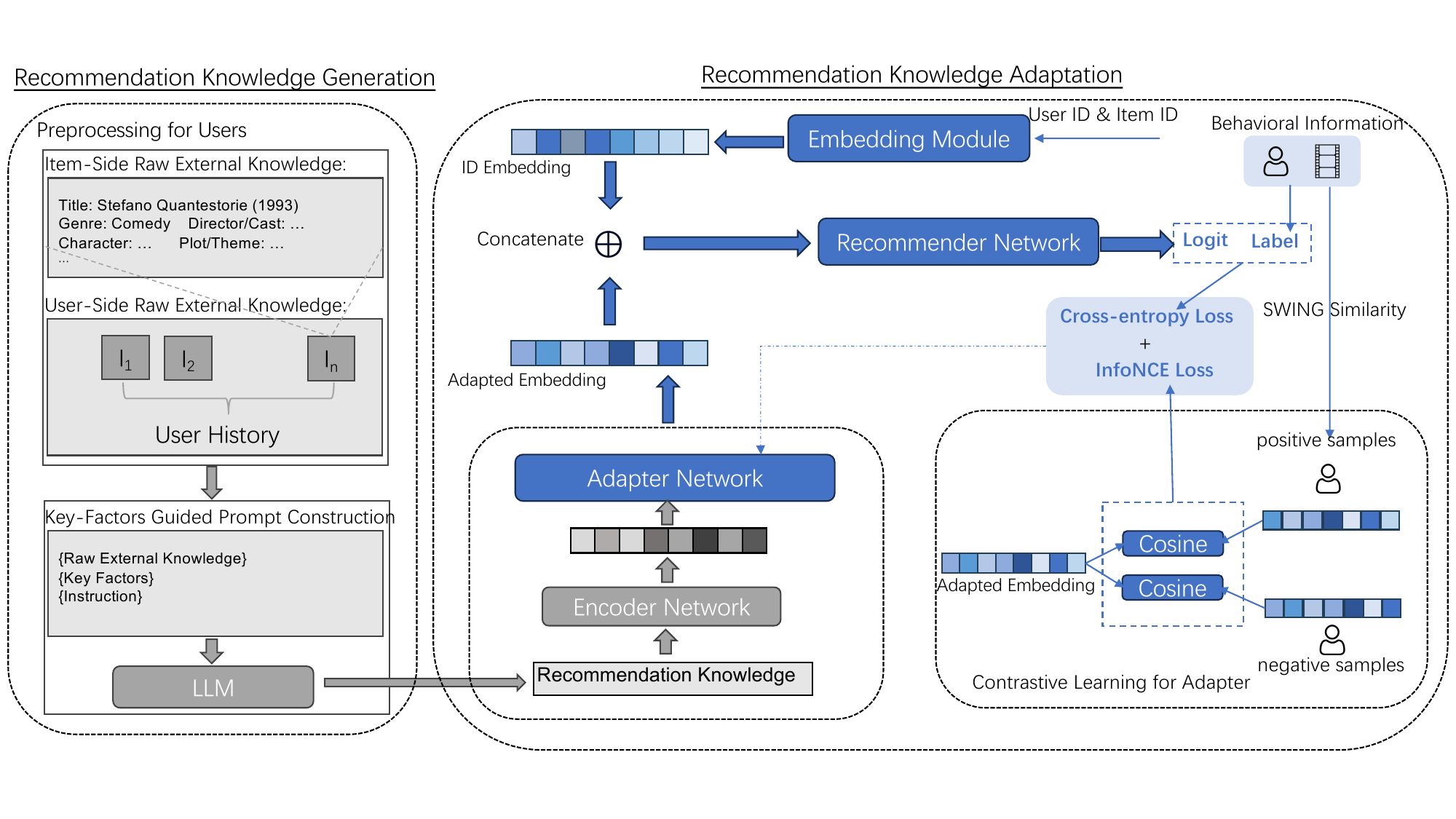}
\vspace{-0.9cm}
\caption{Method Framework}
\label{fig:method-framework}
\end{figure*}

\subsection{Framework Overview}
\label{subsec:method-framework-overview}

The overall structure of TRAWL is illustrated in Fig~\ref{fig:method-framework}, comprising two primary modules: recommendation knowledge generation, and recommendation knowledge adaptation,.

\subsubsection{Recommendation Knowledge Generation}
This module is responsible for extracting recommendation knowledge from raw external sources, involving preprocessing procedures and key-factor guided prompt construction. These processes are detailed in Section~\ref{subsubsec:method-preprocess} and Section~\ref{subsubsec:method-key-factor-construct}.

\subsubsection{Recommendation Knowledge Adaptation}

This module encompasses three types of neural network. An encoder network encodes the recommendation knowledge generated by the LLM into semantic embeddings. An adapter network then transforms these semantic embeddings from the semantic embedding space to the recommendation embedding space. These transformed embeddings are concatenated with ID embeddings and input into the recommender network to produce the final recommendation results. 

TRAWL utilizes a parameter-efficient training methodology, which involves freezing the parameters of the encoder network while jointly training the adapter network and the recommender network. Specifically, the adapter network employs a contrastive learning-based training strategy to integrate behavioral information with semantic information.
Further details are provided in Section~\ref{subsec:method-klg-adaptation-injection}.

\subsection{Recommendation Knowledge Generation}

\subsubsection{Preliminaries}
Prior to detailing the specific procedures of preprocessing and prompt design, we first outline the fundamental assumptions TRAWL makes. In this study, we assume that external knowledge is available in the form of unstructured free text. For instance, for a given item $i$, its associated external knowledge $klg_i$ might be a Wikipedia article that introduces a concept closely related to $i$. On the user side, the external knowledge is less clearly defined compared to that of the items. For a particular user $u$, we assume the recommender system maintains the interaction history of $u$, denoted as $hist_u$. If available, the profile information of $u$ is represented as $pf_u$. The aforementioned textual information $pf_u$, $hist_u$ and $klg_i$ serves as a component to construct the prompts.

\subsubsection{Raw External Knowledge Preprocessing for Users}
\label{subsubsec:method-preprocess}

As mentioned in Section~\ref{Sec:Intro}, we leverage a large language model to extract recommendation knowledge for both users and items. For users, the external knowledge is implicitly represented by the external knowledge of items with which the user has interacted. Given that the LLM processes text-based input, we construct the user's raw external knowledge as follows: For a particular user $u$, we select the most recent $k$ items that have interacted with $u$, denoted as ${i_1, i_2, ..., i_n}$, indexed in ascending order by timestamp. The corresponding raw external knowledge for these items is ${klg_1, klg_2, ... klg_k}$. The raw external knowledge $klg_u$ for $u$ is then formulated as $klg_1 \oplus klg_2 \oplus ... \oplus klg_k$, where $\oplus$ represents the text concatenation operation. In other words, the user's raw external knowledge is preprocessed as the concatenation of the raw external knowledge of the recent $k$ items with which the user has interacted.

\subsubsection{Key-Factors Guided Prompt Construction}
\label{subsubsec:method-key-factor-construct}

We employ a key-factor guided strategy to direct the LLM in extracting valuable recommendation knowledge from raw external knowledge. Key factors represent the most critical elements influencing recommendation quality from a semantic perspective, which are indicated by experts. For instance, in movie recommender systems, key factors may include genres, themes, and awards. By incorporating key factors for both users and items, we enable the LLM to generate high-quality and consistent recommendation knowledge, which enhances the performance of the downstream recommender network.

Subsequently, we integrate the previously mentioned raw external knowledge, key factors, and necessary instructions to finalize the prompt construction for the item side. For the user side, in addition to the same configuration as the item's, user profile information will also be included if available.

\begin{figure*}
\centering
\includegraphics[width=16cm]{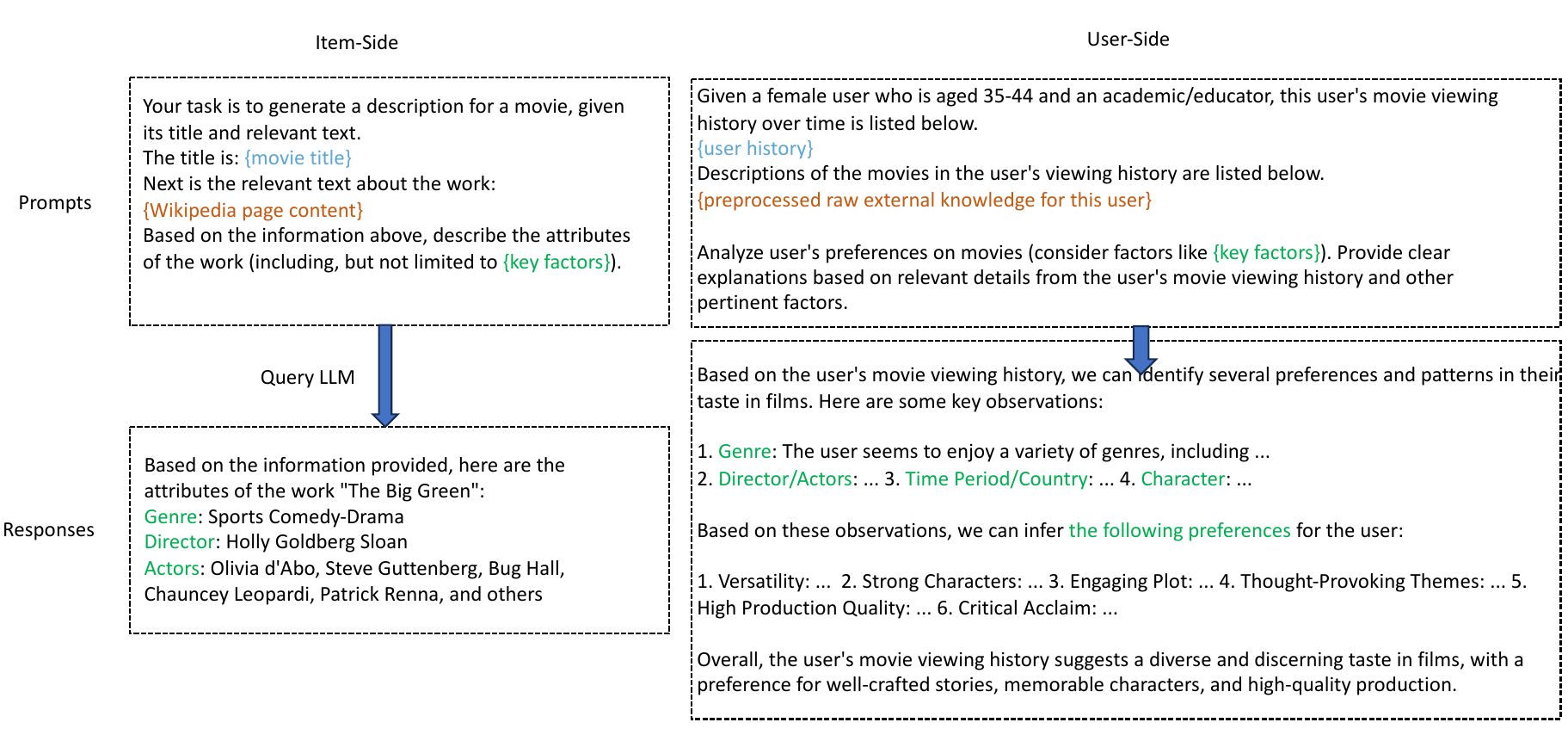}
\vspace{-0.4cm}
\caption{An Illustrative example of calling an LLM for movie recommender network.}
\label{fig:method-example}
\end{figure*}

The constructed prompts are fed into the LLM to generate high-quality recommendation knowledge for both users and items, which will be utilized in the knowledge adaptation and injection phase of TRAWL. Figure~\ref{fig:method-example} presents a prompt instance for a movie recommender system.

\subsection{Recommendation Knowledge Adaptation}
\label{subsec:method-klg-adaptation-injection}

In this module, TRAWL converts the generated recommendation knowledge for users and items into embedding representations, which are then integrated with the backbone recommender network to produce recommendation results. This process involves three types of networks: the encoder network, the adapter network, and the recommender network.

As a first step, TRAWL utilizes an encoder network (i.e., a language model) to encode the recommendation knowledge into text embeddings.
These text embeddings reside in a semantic space, which is suboptimal for direct use by the downstream recommender network.
As discussed in Section~\ref{Sec:Intro}, the semantic representations of recommendation knowledge generated by the LLM exhibit a natural disparity compared to the behavioral representations in the recommender network, which are trained to learning the collaborative relations between users and items. TRAWL employs a Mixture of Experts (MoE) network as an adapter (the training strategy for which is detailed in Section~\ref{subsec:method-klg-adaptation-injection}) to inject behavioral supervision signals into the semantic representation as depicted in Figure~\ref{fig:method-framework}.

It is noteworthy that TRAWL does not impose strict constraints on the specific recommender network, allowing flexibility according to the application context. For simplicity, we assume that an original version of the recommender network (i.e., a recommender network independent of the TRAWL framework) utilizes ID feature embeddings as inputs. TRAWL concatenates the ID embeddings with the adapted semantic embeddings, creating an augmented embedding that combines both semantic and behavioral information. This augmented embedding replaces the original ID embedding as input to the recommender network.

In order to enhance the training of the adapter network, thereby facilitating a more effective transformation of semantic embeddings to the recommendation task, TRAWL employs a behavioral information supervised contrastive learning task for adapter network training. We first explain the method of positive sampling, followed by the design of the loss function.

\textbf{Behavioral Information Based Positive Sampling.}
The objective of the contrastive learning task is to incorporate behavioral information into the semantic representation, ensuring that the representations of two users (or two items) with similar interaction histories are similar. TRAWL applies contrastive learning-based training to both the user and item sides. Below, we provide a detailed explanation of the positive sampling process for users, noting that the positive sampling process for items is analogous.

\begin{equation}
\label{eq:swing-user}
s(u, v) = \Sigma_{i \in I_u \cap I_v}\Sigma_{j \in I_u \cap I_v} \frac{1}{\alpha + \| U_i \cap U_j \|}
\end{equation}

TRAWL employs SWING similarity~\cite{yang2020large_swing} as a metric to calculate the behavioral similarity between two users (or items). Equation~\ref{eq:swing-user} defines this calculation for users (and similarly for items), where $u$ and $v$ denote two specific users, $I_u$ and $I_v$ represent the sets of items interacted with by $u$ and $v$, respectively, $U_i$ and $U_j$ represent the sets of users that have interacted with items $i$ and $j$, respectively, and $\alpha$ iss a smoothing term. The intuitive explanation of SWING similarity is that if users $u$ and $v$ have interacted with items $i$ and $j$, and items $i$ and $j$ have a comparatively small overlap in their interacting users, then $u$ and $v$ are likely to have similar preferences. By considering the overlapping users, SWING similarity mitigates the influence of super-items (items with a high number of interactions) and provides a more accurate estimation compared to the simple Jaccard similarity coefficient~\cite{jaccard1901etude, yang2020large_swing}.

Using SWING similarity as a metric, TRAWL can rank users based on behavioral information for a given user. Specifically, for each user $u$ in the user-item interaction training data, TRAWL selects the user with the highest SWING similarity from the remaining users as the positive sample. During training, the adapter network aims to reduce the distance between the embeddings of positive pairs, thereby integrating behavioral information into the original semantic representations.

\textbf{Contrastive Loss Design.}
Given the positive samples identified using SWING similarity, TRAWL employs the infoNCE loss~\cite{gutmann2010noise, oord2018representation} as the contrastive learning training objective, formulated by Equation~\ref{eq:infoNCE-user} for the user side (the item-side loss is defined analogously). 

In this context, $B$ represents the training batch size, $\mathbf{u}_j$ denotes the normalized embedding of the $j$-th user in the batch produced by the adapter network, and $\mathbf{u}_j^p$ denotes the normalized adapted embedding of the positive sample corresponding to $\mathbf{u}_j$. The temperature coefficient is denoted by $\tau$.

\begin{equation}
\label{eq:infoNCE-user}
\ell_{uu} = -\sum\limits_{j=1}^{B}\log\frac{\exp(\mathbf{u}_j\cdot\mathbf{u}^p_j/\tau)}{\exp(\sum_{j'=1}^B\mathbf{u}_{j'}\cdot\mathbf{u}^p_{j'}/\tau)}
\end{equation}

\subsubsection{Parameter-efficient Multi-task Joint Training}
\label{subsubsec-method-pe-ml-joint-training}

During training, TRAWL freeze the parameters of the encoder network and only adjust the parameters of the adapter network and the recommender network, thereby achieving parameter-efficient training. Given that the recommender network and the adapter network have distinct training objectives, we implement a multi-task joint training approach. The final loss $\mathcal{L}$ is formulated in Equation~\ref{eq:sum-loss}, where $L_{rec}$ denotes the loss of the recommender network. The weights $w_1$ and $w_2$ are tuneable hyperparameters. During the training process, the adapter network and the recommender network are optimized simultaneously.

\begin{equation}
\label{eq:sum-loss}
\mathcal{L} = L_{rec} + w_1 \cdot L_{uu} + w_2 \cdot L_{ii}
\end{equation}

\section{Experiment Setup}
\label{sec:exp-setup}

This section outlines the experimental setup and configuration.

\subsection{Dataset and Baseline Methods}

\textbf{Dataset} For the public dataset experiment, we utilize the MovieLens-1M dataset. The \textit{MovieLens-1M}\footnote{https://grouplens.org/datasets/movielens/1m/} dataset is extensively employed in academic research on movie recommendation systems. In this dataset, users provide explicit ratings ranging from 1 to 5 for movies, where a rating of 5 indicates a strong preference for the movie, and a rating of 1 indicates strong disfavor. The dataset comprises 6,640 users, 3,883 movies, and over 1 million user-movie interactions.

\textbf{Baselines} As introduced in Section \ref{Sec:Intro}, TRAWL employs large language models (LLMs) to extract recommendation knowledge from raw external knowledge sources. To evaluate its necessity, we compare TRAWL with three types of baselines. The first type utilizes raw external knowledge without any denoising operations. The second type includes summary-based models, which are further classified into basic summarization and topic-focused summarization models. The third type comprises LLM-generated external knowledge, derived from the knowledge contained within LLMs.

For the basic summarization model, we use \textbf{PEGASUS}~\cite{zhang2020pegasus}, which achieves high-quality text summarization by training a Transformer model with a self-supervised pretraining objective. For topic-focused summarization, we employ \textbf{CTRLsum}~\cite{he2020ctrlsum}, which introduces control tokens during training and inference, allowing users to interact with the summarization system using keywords or descriptive prompts. Both models achieve state-of-the-art performance among models with a similar parameter magnitude.

For LLM-generated external knowledge, we adopt the methodology described in \textbf{KAR}~\cite{xi2023towards} to generate external knowledge from the knowledge embedded in LLMs.

TRAWL is designed to be versatile and compatible with various backbone recommender networks. To validate this flexibility, we select a group of widely-used click-through rate (CTR) prediction networks as the backbone recommender networks in TRAWL. These include DeepFM~\cite{guo2017deepfm}, DIEN~\cite{zhou2019deep}, and DIN~\cite{zhou2018deep}.

\subsection{Task and Metrics}

We evaluate the effectiveness of TRAWL using the \textit{click-through rate (CTR) prediction task}, which is a fundamental task for assessing recommender system performance. The primary objective of the CTR prediction task is to estimate the probability that a user will click on an item. We employ \textit{AUC} (Area Under the Curve) and \textit{LogLoss} (Logarithmic Loss) as our evaluation metrics. Specifically, for the ML-1M dataset, we classify rating scores greater than 3 as positive samples (label 1) and those less than or equal to 3 as negative samples (label 0).

\subsection{Other Details}
\label{subsubsec:exp-setup-other-details}
\textbf{Raw External knowledge Acquisition} We collected the raw external knowledge for the movies in the MovieLens-1M dataset from Wikipedia. Specifically, we downloaded a Wikipedia dump\footnote{https://www.kaggle.com/kenshoresearch/kensho-derived-wikimedia-data} as our global knowledge base and filtered out all movie-type entities based on WikiProject Movies\footnote{https://www.wikidata.org/wiki/Wikidata:WikiProject\_Movies/Properties}. For each movie in the dataset, we identified the corresponding Wikipedia entity by calculating the edit distance similarity between the movie title and the entity label. Finally, we extracted the contents of the Wikipedia page corresponding to each movie entity as the raw external knowledge.

\textbf{LLM} We utilized a locally-deployed LLM\footnote{https://huggingface.co/internlm/internlm2-7b} in our experiments.

\textbf{Knowledge encoder} We employed the bert-base-uncased\footnote{https://huggingface.co/bert-base-uncased} model for the MovieLens-1M dataset.

\textbf{Training details} We partitioned the users into training, validation, and test sets in an 8:1:1 ratio. For TRAWL's hyper-parameters, we conducted a parameter sweep on the contrastive learning loss weights as defined in Equation~\ref{eq:sum-loss} and the temperature parameters in Equations~\ref{eq:infoNCE-user}. Detailed sensitivity analyses are presented in Section~\ref{subsubsec:para-sensitivity}. For training-specific hyper-parameters, we fine-tuned the batch size ($256$, $512$, $1024$) and learning rate ($1e-4$, $5e-4$). The final experimental results were achieved with a batch size of $256$, a learning rate of $1e-4$, a temperature of $0.15$, and contrastive learning loss weights of $0.004$ for the user and $0.008$ for the item.

\section{Experiment Results}
\label{sec:exp-results}

This section presents the experimental results on a public dataset and a real-world online recommendation scenario.

\begin{table}[t]
\centering
\caption{Experiment Results for CTR prediction Task on MovieLens-1M (lower LogLoss indicates better results)}
\vspace{-0.3cm}
\centering
\small
\begin{tabular} {|l|l|l|l|}
\hline
Backbone & Knowledge Type & AUC & LogLoss \\
\hline

\multirow{6}*{DIN}  & Raw & 0.769 & 0.5639 \\
\cline{2-4}
 & w/o external knowledge & 0.7796 & 0.5581 \\
\cline{2-4}
 & Basic Summary & 0.7726 & 0.5625 \\
\cline{2-4}
 & Topic-focused Summary & 0.7739 & 0.5602 \\
\cline{2-4}
 & $LLM_{KAR}$ & 0.7794 & 0.5545 \\
\cline{2-4}
 & $LLM_{TRAWL}$ & \textbf{0.7812} & \textbf{0.5538} \\
\hline
\multirow{6}*{DIEN}  & Raw & 0.7526 & 0.5821 \\
\cline{2-4}
 & w/o external knowledge & 0.7765 & 0.5599 \\
\cline{2-4}
 & Basic Summary & 0.7558 & 0.5774 \\
\cline{2-4}
 & Topic-focused Summary & 0.7571 & 0.5748 \\
\cline{2-4}
 & $LLM_{KAR}$ & 0.7776 & 0.5589 \\
\cline{2-4}
 & $LLM_{TRAWL}$ & \textbf{0.7789} & \textbf{0.5572} \\
\hline
\multirow{6}*{DeepFM}  & Raw & 0.7672 & 0.5665 \\
\cline{2-4}
 & w/o external knowledge & 0.7788 & 0.5556 \\
\cline{2-4}
 & Basic Summary & 0.7689 & 0.5635 \\
\cline{2-4}
 & Topic-focused Summary & 0.7702 & 0.5611 \\
\cline{2-4}
 & $LLM_{KAR}$ & 0.7787 & 0.5572 \\
\cline{2-4}
 & $LLM_{TRAWL}$ & \textbf{0.7792} & \textbf{0.5550} \\
\hline

\end{tabular}
\label{Tab:Exp-CTR}
\end{table}

\begin{table}[t]
\centering
\caption{Ablation Experiment Results for CTR Prediction.}
\vspace{-0.3cm}
\centering
\small
\begin{tabular} {|l|l|l|l|l|}
\hline
Backbone & Metric & TRAWL & w/o CL Loss & w/o Adapter\\

\hline
\multirow{2}*{DIN} & AUC & \textbf{0.7812} & 0.7767~(-0.58\%) & 0.7749~(-0.81\%) \\
\cline{2-5}
 & LogLoss & \textbf{0.5538} & 0.5563~(+0.45\%) & 0.5582~(+0.79\% \\
\hline
\multirow{2}*{DIEN} & AUC & \textbf{0.7789} & 0.7754~(-0.45\%) & 0.7737~(-0.67\%) \\
\cline{2-5}
 & LogLoss & \textbf{0.5572} & 0.5621~(+0.88\%) & 0.5645~(+1.31\% \\
\hline
\multirow{2}*{DeepFM} & AUC & \textbf{0.7792} & 0.7750~(-0.54\%) & 0.7732~(-0.77\%) \\
\cline{2-5}
 & LogLoss & \textbf{0.5550} & 0.5589~(+0.7\%) & 0.5603~(+0.95\% \\
\hline

\end{tabular}
\label{Tab:Exp-ablation}
\end{table}

\subsection{Main Results}

Table~\ref{Tab:Exp-CTR} presents the experimental results of four baselines combined with three backbone recommender networks on the MovieLens-1M dataset. Here, $LLM_{TRAWL}$ represents recommendation knowledge extracted from raw external knowledge by TRAWL, and $LLM_{KAR}$ represents recommendation knowledge generated directly by an LLM. From the experimental results, we can derive five key observations:

\textbf{Effectiveness of External Knowledge.} Comparing the results without external knowledge to those with external knowledge, we observe a significant improvement, indicating that integrating external knowledge into the recommender network enhances performance.

\textbf{Basic Summarization Performance.} The basic summarization model performs worse than using raw external knowledge. This suggests that basic summarization models may overlook valuable information beneficial to recommendations, thereby failing to effectively extract recommendation knowledge from raw external knowledge.

\textbf{Topic-Focused Summarization Performance} The topic-focused summarization model achieves better results than the basic summarization model but still lags behind raw external knowledge. This indicates that focusing on recommendation-related topics helps produce better summaries for recommender networks, but the lack of common-sense inference means it may still miss valuable information for recommendations.

\textbf{LLM-Generated External Knowledge.} Using external knowledge generated directly by an LLM yields results comparable to raw external knowledge but is still inferior to TRAWL. This suggests that while the knowledge stored in LLMs is beneficial for recommendation systems, the improvement is limited, possibly due to hallucination issues.

\textbf{TRAWL Performance} TRAWL achieves the best performance across all backbone recommender networks, indicating that it outperforms simple summarization by successfully extracting recommendation knowledge from raw external knowledge. The consistent improvement across all backbone recommender networks validates the versatility of TRAWL.
\emph{In summary, these observations collectively suggest that external knowledge has significant potential to enhance recommendation performance. However, simple summarization may result in suboptimal outcomes by missing valuable aspects of raw external knowledge. In contrast, LLMs, capable of both summarization and common-sense inference, can effectively extract recommendation knowledge when appropriately guided by prompts.}

\vspace{-0.3cm}
\subsection{Effectiveness of Adaptive Training}

To validate the necessity of the adapter network and the effectiveness of the contrastive learning training strategy, we conducted an ablation study with two levels of ablation, as presented in Table~\ref{Tab:Exp-ablation}. This study yields two key findings.
The first finding indicates that when the auxiliary contrastive loss is removed, and the adapter network and recommender network are trained solely with the recommendation loss, the recommendation performance considerably degrades across all three backbone recommender networks.
The second observation reveals that removing the entire adapter network and directly integrating the original semantic embedding into the recommender network results in a further decline in recommendation performance.

These findings validate that \emph{the adapter network is crucial for bridging the gap between semantic space and recommendation space. Additionally, the contrastive learning approach effectively integrates behavioral information with semantic information, leading to enhanced performance.}

\vspace{-0.2cm}
\subsection{Parameter Sensitivity}
\label{subsubsec:para-sensitivity}

\begin{figure}[htbp]
    \centering
    \subfigure[User CL Loss Weight]{
        \includegraphics[width=0.3\columnwidth]{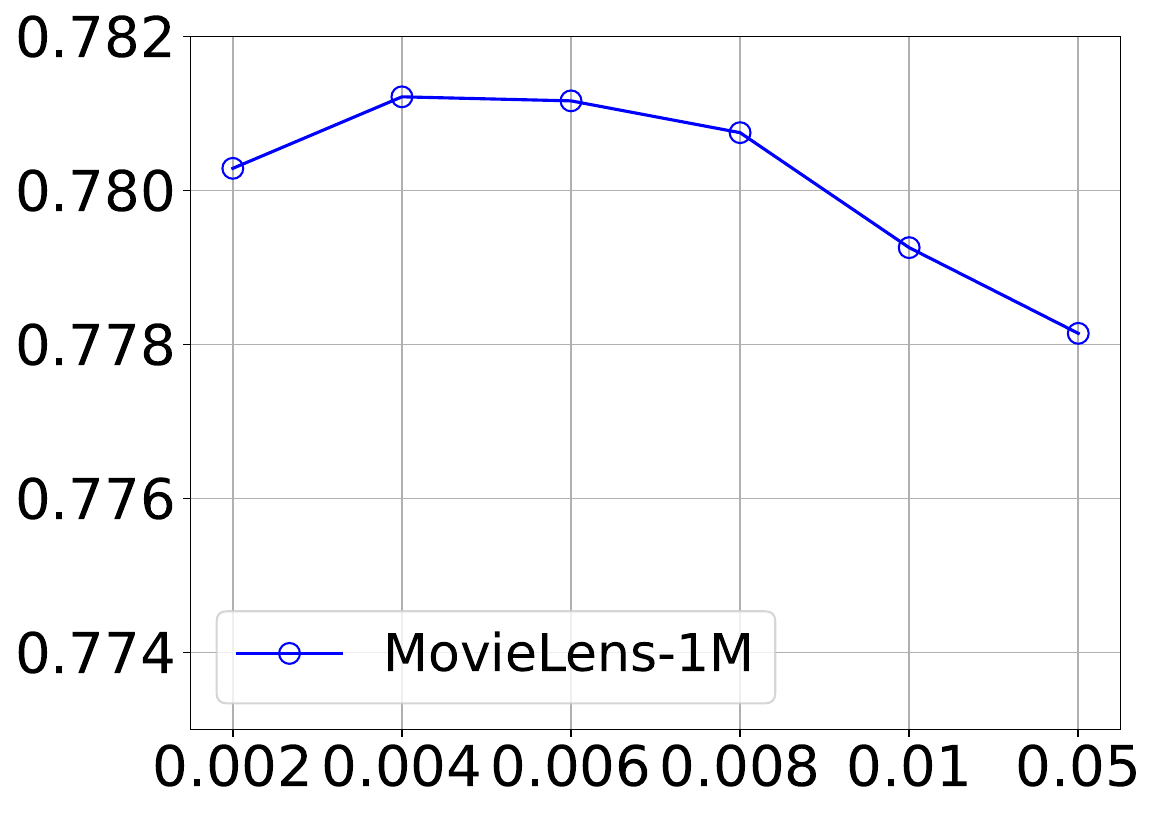}
    }
    \subfigure[Item CL Loss Weight]{
        \includegraphics[width=0.3\columnwidth]{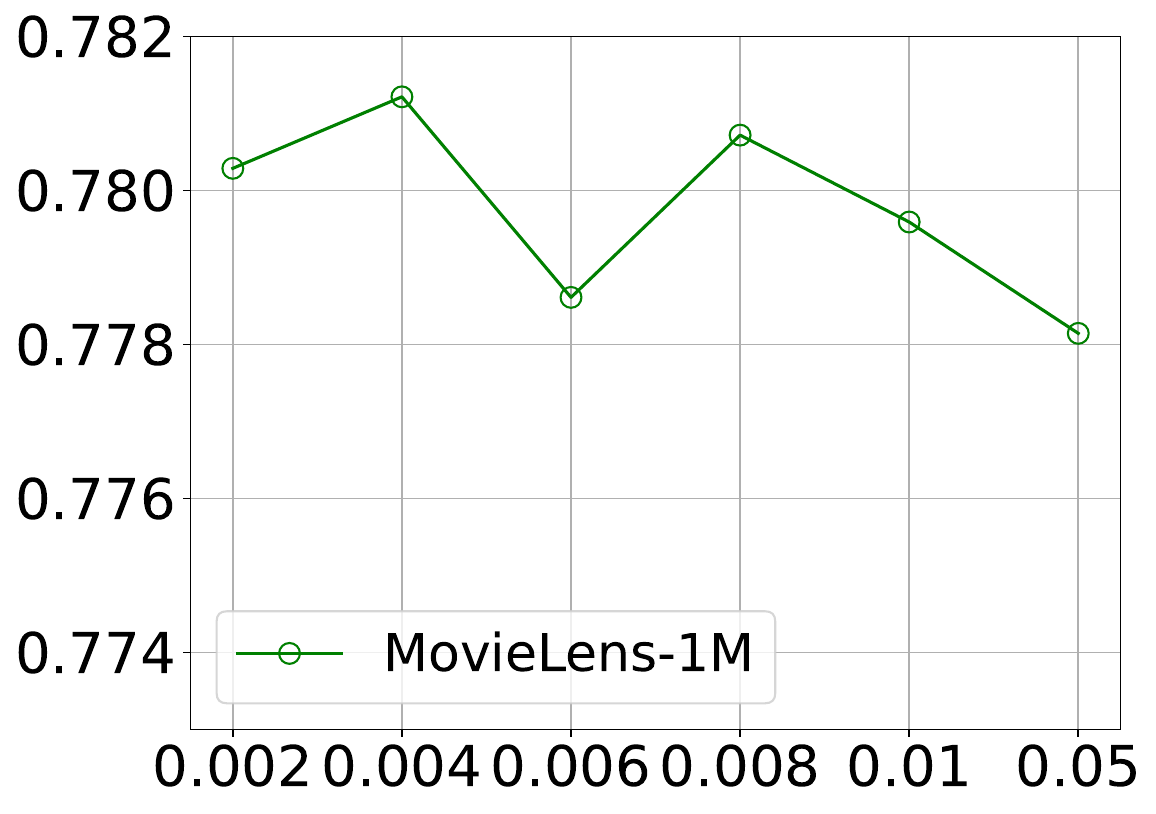}
    }
    \subfigure[Temperature]{
        \includegraphics[width=0.3\columnwidth]{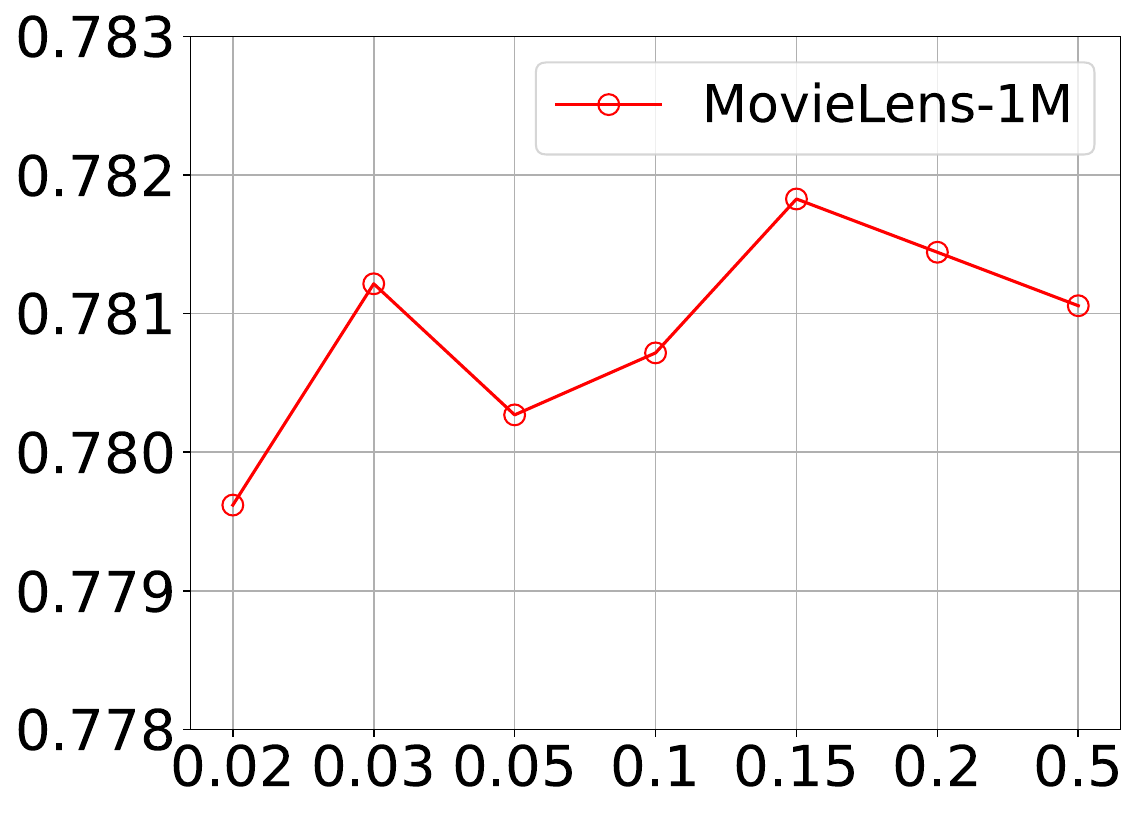}
    }
\vspace{-0.5cm}
    \caption{Hyper-parameter}
    \label{fig:hyper-param}
\end{figure}

This section examines the sensitivity of hyper-parameters in TRAWL.
Specifically, we analyze the contrastive loss weights $w_1$ and $w_2$ in Equation~\ref{eq:sum-loss} and the temperature $\tau$ in Equations~\ref{eq:infoNCE-user}. To assess the sensitivity of each hyper-parameter, we fixed the others and tuned the parameter within the specified range observing the AUC metric on the validation set. The results are displayed in Figure~\ref{fig:hyper-param}. For the contrastive loss weight on the user side, the optimal value is $0.004$; values lower than this have minimal impact on the adapter, while higher values may have excessive influence. On the item side, similar trends are observed, with the optimal value at $0.008$. For the temperature hyper-parameter $\tau$, the optimal value is $0.15$. A higher temperature diminishes the model's ability to distinguish between positive and negative samples, whereas a lower temperature leads to the opposite effect.
\vspace{-0.2cm}
\subsection{Online Experiment}
\label{subsec-online}

To assess the effectiveness of TRAWL in a real-world environment, we conducted an online experiment using the WeChat Article Recommendation Service, which is the article recommendation service of WeChat, the largest social platform in China. In this service, users receive recommendations for news and articles written by other users. An in-house knowledge base and its associated entity linking tool were utilized to identify raw external knowledge for the articles. The knowledge base comprises 96 million multi-language entities, 11 million Chinese entities, and 690 million relations, sourced from Wikipedia
CN-DBpedia\footnote{http://kw.fudan.edu.cn/cndbpedia/intro/}, and in-house WeChat Subscriptions and Live business data.

Following the methodology outlined in Section~\ref{sec:method}, we utilized Qwen1.5-7B
to extract recommendation knowledge. The bge-large-zh 
model was employed as the encoder network.

For training the adapter network, positive samples were selected based on the SWING similarity calculated from user-item interaction data from the previous day, as described in Section~\ref{subsec:method-klg-adaptation-injection}. During the prediction phase, for items with which the user had previously interacted positively, the most similar items were recalled based on the similarity of the adapted item embeddings and added to the existing item recall pool.

The online A/B test encompassed 8 million users and 1 million articles. The results, presented in Table~\ref{Tab:Exp-ablation}, indicate that three core metrics of the recommendation system—average click count per user, average read seconds per user, and average interaction count per user—increased by 0.310\%, 0.360\%, and 0.543\%, respectively, compared with the baseline online model. This online A/B test validates the effectiveness of TRAWL for real-world online recommender systems.

\begin{table}[t]
\centering
\caption{Online A/B test results. CC: click count. RS: read seconds. IC: interaction count.}
\vspace{-0.2cm}
\centering
\small
\begin{tabular} {|l|l|l|l|}
\hline
Model & Avg.CC & Avg.RS & Avg.IC \\
\hline
Online Model & 2.553 & 168.976 & 0.0618 \\
\hline
TRAWL & 2.561 & 169.584 & 0.0621 \\
\hline
\# improve & $0.310\% \pm 0.224\%$ & $0.360\% \pm 0.302\%$ & $0.543\% \pm 1.093\%$ \\
\hline
\end{tabular}
\label{Tab:Exp-ablation}
\end{table}

\section{Conclusion}
\label{sec:conclusion}

This work proposed TRAWL, an external knowledge-enhanced recommendation framework. By utilizing a large language model to extract useful recommendation knowledge from raw external sources, along with an adapter network employing a contrastive learning-based training strategy, TRAWL effectively addresses two key challenges outlined in Section~\ref{Sec:Intro}: denoising raw external knowledge and adapting semantic representations. The framework's efficacy is validated through extensive experiments on public datasets  well as real-world online recommendation scenarios. Beyond its practical value in enhancing recommendations, TRAWL introduces new possibilities for integrating large language models with recommender systems and combining semantic information with behavioral data.

\bibliographystyle{ACM-Reference-Format}
\bibliography{main}

\end{document}